\documentstyle[12pt]{article}

\newcommand{\be}{\begin{equation}}
\newcommand{\ee}{\end{equation}}
\newcommand{\Dlt}{\Delta}
\newcommand{\dlt}{\delta}

\newcommand{\br}{{\bf r}}
\newcommand{\bS}{{\bf S}}
\newcommand{\bk}{{\bf k}}

\newcommand{\bt}{\beta}
\newcommand{\vp}{\varphi}
\newcommand{\ep}{\varepsilon}
\newcommand{\al}{\alpha}
\newcommand{\ra}{\rightarrow}

\newcommand{\om}{\omega}
\newcommand{\Om}{\Omega}
\newcommand{\Gm}{\Gamma}
\newcommand{\dgr}{\dagger}

\newcommand{\cF}{{\cal F}}

\begin{document}

\begin{center}
{\Large{\bf Bose-Einstein condensation and gauge symmetry
breaking} \\ [5mm]
V.I. Yukalov} \\ [3mm]

{\it Bogolubov Laboratory of Theoretical Physics, \\
Joint Institute for Nuclear Research, Dubna 141980, Russia}

\end{center}

\vskip 2cm

\begin{abstract}

The fundamental problem is analized, the relation between Bose-Einstein
condensation and spontaneous gauge symmetry breaking. This relation is
largerly misunderstood in physics community. Numerous articles and
books contain the statement that, though gauge symmetry breaking helps
for describing Bose-Einstein condensation, but the latter, in principle,
does not require any symmetry breaking. This, however, is not correct.
The analysis is based on the known mathematical theorems. But in order
not to overcomplicate the presentation and to make it accessible to all
readers, technical details are often omitted here. The emphasis is made
on the following basic general facts: Spontaneous breaking of gauge
symmetry is the {\it necessary and sufficient} condition for Bose-Einstein
condensation. Condensate fluctuations, in thermodynamic limit, are {\it
negligible}. Their catastrophic behavior can arise only as a result of
incorrect calculations, when a Bose-condensed system is described without
gauge symmetry breaking. It is crucially important to employ the {\it
representative statistical ensembles} equipped with all conditions that
are necessary for a unique and mathematically correct description of
the given statistical system. Only then one is able to develop a 
self-consistent theory, free of paradoxes.

\end{abstract}

\vskip 1cm


\vskip 1cm


\newpage

\section{Introduction}

In recent years, the phenomenon of Bose-Einstein condensation (BEC) has
received much attention, both experimentally and theoretically (see, e.g.,
the book [1] and review articles [2--6]). At the present time, Bose-Einstein
condensates of trapped atoms have been realized in about 70 laboratories of
15 countries, USA, Germany, France, New Zealand, England, Japan, Italy,
Netherlands, Israel, Australia, China, Austria, Switzerland, Canada, 
and Brazil. Magnetic, magneto-optical, and all-optical traps have been 
employed. BEC has been achieved for 11 atomic species ($^1$H, $^4$He, 
$^7$Li, $^{23}$Na, $^{39}$K, $^{41}$K, $^{52}$Cr, $^{85}$Rb, $^{87}$Rb, 
$^{133}$Cs, and $^{147}$Yb), four types of molecules formed by Bose atoms 
($^{23}$Na$_2$, $^{85}$Rb$_2$, $^{87}$Rb$_2$, and $^{133}$Cs$_2$), and for 
two types of molecules formed by Fermi atoms ($^{6}$Li$_2$ and 
$^{40}$K$_2$). More details can be found in Refs. [2,5,7--10].

One of course, always remembers that BEC is assumed to exist in superfluid 
$^4$He, which can be revealed, e.g., in scattering experiments with neutrons 
[11--14] and $x$-rays [15,16]. Also, BEC of boson-type quark clusters might 
arise in interiors of neutron stars and in heavy-ion collisions [17--19]. 
There exist several other examples of BEC, which will be cited in the last 
section of the paper. Thus, Bose-condensed systems form quite a numerous 
family.

In the theory of Bose-condensed systems, there have remained some
principal problems that have not been properly understood in a large part
of physics community. One of such fundamental problems is the relation
between BEC and the U(1) gauge symmetry breaking. To understand the issue,
it is necessary to give answers to the following group of mutually
interrelated questions: Is gauge symmetry breaking only a sufficient
condition for BEC or it is also a necessary one? In what sense one should
interpret the statement that BEC and gauge symmetry breaking are equivalent?
What is the difference between the two methods of gauge symmetry breaking,
with the help of infinitesimal sources and by means of the Bogolubov shift?
What is the origin of the catastrophic behavior of condensate fluctuations
in Bose-condensed systems? And what are the requirements for making a theory,
with broken gauge symmetry, self-consistent? The aim of the present paper 
is to answer these questions.

\section{Expansions over natural orbitals}

We shall consider a statistical system of $N$ bosons in volume $V$. For
simplicity, we keep in mind the particles with no internal degrees of
freedom. After answers to the principal questions are understood, it is
easy to generalize the consideration to particles with internal degrees
of freedom, when fragmented condensates [20] might arise.

The basic variable, characterizing particles of a statistical system, is
the field operator $\psi(\br,t)$, where $\br$ is spatial vector and $t$,
time. Bose particles correspond to the Bose commutation relations
\be
\label{1}
[\psi(\br,t),\; \psi^\dgr(\br',t)] = \dlt(\br-\br') \; ,
\ee
with other commutation relations for $[\psi,\psi]$ and $[\psi^\dgr,\psi^\dgr]$
being zeros.

\subsection{Expansion of density matrix}

A general way for characterizing BEC is the Penrose-Onsager scheme [21],
based on the consideration of the eigenvalues of the first-order density
matrix, which is defined as the statistical average
\be
\label{2}
\rho(\br,\br',t) \; \equiv \;
< \psi^\dgr(\br',t) \psi(\br,t) > \; .
\ee
The eigenvalues $n_k$ are given by the eigenproblem
\be
\label{3}
\int \rho(\br,\br',t) \; \vp_k(\br',t) \; d\br' =
n_k(t)\; \vp_k(\br,t) \; ,
\ee
whose eigenfunctions $\vp_k$ are termed {\it natural orbitals} [22]. The
density matrix (2) acquires the spectral resolution
\be
\label{4}
\rho(\br,\br',t) = \sum_k n_k(t) \; \vp_k(\br,t) \; \vp_k^*(\br',t)
\ee
over the natural orbitals.

For an equilibrium system, the density matrix (2) does not depend on
time. Respectively, $n_k$ and $\vp_k$ do not depend on time as well, so
that the spectral resolution (4) becomes
\be
\label{5}
\rho(\br,\br') = \sum_k n_k \vp_k(\br) \vp_k^*(\br') \; .
\ee
In Eqs. (3), (4), and (5), the index $k$ is an appropriate multi-index
labelling the eigenvalues and eigenfunctions of eigenproblem (3).

In what follows, we shall mainly deal with the equilibrium case (5),
keeping in mind that it can straightforwardly be extended to the general
nonequilibrium situation of Eq. (4).

Among the eigenvalues $n_k$, let is separate the largest one
\be
\label{6}
N_0 \equiv \sup_k n_k \; ,
\ee
related to a state $\vp_0(\br)\equiv\vp_{k_0}(\br)$ labelled by a
multi-index $k_0$. One says that there occurs BEC, when the largest
eigenvalue $N_0$ is proportional to the total number of particles $N$,
so that $N_0\propto N$ for all $N$, including the thermodynamic limit,
when
\be
\label{7}
N \ra \infty \; , \qquad V \ra \infty \; , \qquad
\frac{N}{V} \ra const \; ,
\ee
where a positive constant is implied. Thence, BEC, by definition, is
characterized by the condition
\be
\label{8}
\lim_{N\ra\infty} \; \frac{N_0}{N} >  0\; ,
\ee
where the limit means the thermodynamic limit (7). Under this condition,
the spectral resolution (5) can be rewritten as the sum
\be
\label{9}
\rho(\br,\br') = N_0 \vp_0(\br) \vp_0^*(\br') +
\sum_{k\neq 0} n_k \vp_k(\br) \vp_k^*(\br') \; .
\ee
The natural orbitals are assumed to be normalized to one, $||\vp_k||=1$.
Therefore, $N_0$ is the number of condensed particles.

Often one connects the appearance of BEC with the off-diagonal long-range
order, defining the condensate density as the limit
$$
\rho_0 \equiv \frac{N_0}{V} =\lim_{r\ra\infty} \rho(\br,0) \; .
$$
This, however, has sense only for an equilibrium uniform system, when
$\vp_0=1/\sqrt{V}$ and the second term in Eq. (9) tends to zero with
$r\equiv|\br|\ra\infty$. For example, for a nonuniform confined system,
one has $\vp_k(\br)\ra 0$ as $r\ra\infty$, which would mean that for
this nonuniform system BEC would be impossible [2].

The Penrose-Onsager criterion of BEC, in the form of condition (8),
where $N_0$ is the largest eigenvalue (6), is, clearly, more general.
Another general form of the BEC criterion can be given by introducing the
order indices [22,23]. Let us consider $\hat\rho_1\equiv[\rho(\br,\br')]$
as a matrix with respect to the variables $\br$ and $\br'$. Then the order
index of the matrix is defined [22,23] as
\be
\label{10}
\om(\hat\rho_1) \equiv
\frac{\log||\hat\rho_1||}{\log{\rm Tr}\hat\rho_1} \; ,
\ee
where $||\hat\rho_1||$ is the norm of $\hat\rho_1$. For large $N\gg 1$, 
this takes the form
\be
\label{11}
\om(\hat\rho_1) = \frac{\log N_0}{\log N} \; .
\ee
The order indices (10) or (11) are defined for an arbitrary system and
describe different types of ordering:
$$
\om(\hat\rho_1) < 0 \qquad ({\rm no \; order}) \; ,
$$
$$
\om(\hat\rho_1) =0 \qquad ({\rm short-range \; order}) \; ,
$$
$$
0 < \om(\hat\rho_1) <1 \qquad ({\rm mid-range \; order}) \; ,
$$
$$
\om(\hat\rho_1) = 1 \qquad ({\rm long-range \; order}) \; .
$$
The long-range order happens in thermodynamic limit, when $N\ra\infty$
and $N_0\propto N$. Then, from Eq. (11) it follows that

\be
\label{12}
\lim_{N\ra\infty} \om(\hat\rho_1) = 1 \; .
\ee

The criteria (8) and (12) are, of course, absolutely general and can
be applied to any system. However, being defined through the density
matrix (2), they are not useful without knowing the latter. But the
calculation of the density matrix for an interacting system is not a
trivial task. A practical calculational procedure is to be developed.

\subsection{Expansion of field operators}

Let us specify how the spectral resolution (9) for the density matrix
can be obtained. The field operator $\psi(\br)$ can be expanded over
the basis of natural orbitals,
\be
\label{13}
\psi(\br) = \sum_k a_k \vp_k(\br) \; .
\ee
The operators $a_k$ in the $k$-representation obey the commutation
relations
\be
\label{14}
[a_k,\; a_p^\dgr] = \dlt_{kp} \; , \qquad
[a_k,\; a_p] = 0 \; .
\ee

Generally, we do not know which of the orbitals $\vp_k(\br)$
corresponds to the condensate. The answer to this question is easy
only for the uniform system, when the multi-index $k$ translates into
the momentum $\bk$, so that the condensate is related to $\bk=0$.
But in the general case, we cannot decide which of the orbitals is
that of the condensate. To decide this, we need to know the eigenvalues
of the density matrix (2), with the largest eigenvalue pointing at the
condensate. So that, actually, we are yet in the vicious circle.
Expansion (13), however, is useful for proving some theorems of the
condensate existence.

Assume that we are aware which of the natural orbitals $\vp_k(\br)$
is related to the condesate, and let us mark it as $\vp_0(\br)$.
Then we may separate in expansion (13) the condensate field operator
\be
\label{15}
\psi_0(\br) \equiv a_0 \vp_0(\br)
\ee
from the remaining part associated with the field operator of
uncondensed particles,
\be
\label{16}
\psi_1(\br) \equiv \sum_{k\neq 0} a_k \vp_k(\br) \; .
\ee
Hence, expansion (13) can be represented as the sum
\be
\label{17}
\psi(\br) = \psi_0(\br) +\psi_1(\br) \; .
\ee
The natural orbitals are supposed to be orthonormal,
$$
\int \vp_k^*(\br) \vp_p(\br)\; d\br = \dlt_{kp} \; .
$$
From here, the field operators of condensed and uncondensed particles,
by construction, are orthogonal,
\be
\label{18}
\int \psi_0^\dgr(\br) \psi_1(\br)\; d\br = 0 \; .
\ee

Correctly speaking, the operators $\psi_0(\br)$ and $\psi_1(\br)$
are not separate independent operators, but they are just two parts
of one field operator $\psi(\br)$. All of these operators act on the
Fock space $\cF(\psi)$ generated by the operator $\psi^\dgr$. All
operators from the algebra of observables, being functionals of
$\psi$, are defined on the Fock space $\cF(\psi)$. See mathematical
details in Ref. [24].

Neither of the operators $\psi_0(\br)$ or $\psi_1(\br)$ represents
bosons, since their commutation relations are not of Bose statistics.
Really, the former operators obey the commutation relation
\be
\label{19}
[\psi_0(\br),\; \psi_0^\dgr(\br')] =
\vp_0(\br) \vp_0^*(\br') \; ,
\ee
while the latter, the relation
\be
\label{20}
[\psi_1(\br),\; \psi_1^\dgr(\br')] = \sum_{k\neq 0}
\vp_k(\br) \vp_k^*(\br') \; ,
\ee
other commutation relations, such as $[\psi_0,\psi_0]$,
$[\psi_1,\psi_1]$, and $[\psi_0,\psi_1]$ being zeros. Involving the
$\dlt$-function expansion
$$
\dlt(\br-\br') = \sum_k \vp_k(\br) \vp_k^*(\br') \; ,
$$
that is, the completeness of the natural orbital basis $\{\vp_k(\br)\}$,
the commutation relation (20) can be represented as
\be
\label{21}
[\psi_1(\br),\; \psi_1^\dgr(\br')] = \dlt(\br-\br') -
[\psi_0(\br),\; \psi_0^\dgr(\br')] = \dlt(\br-\br') -
\vp_0(\br) \vp_0^*(\br') \; .
\ee
From the latter equation it follows
$$
\int [\psi_1(\br),\; \psi_1^\dgr(\br')] \psi_0(\br')\; d\br' =
0 \; ,
$$
in agreement with the orthogonality condition (18).

The commutation relations (19), (20), and (21) simplify in thermodynamic 
limit, provided that the condensate natural orbital satisfies the inequality
\be
\label{22}
|\vp_0(\br)|^2 \leq \frac{const}{N^\nu} \qquad (\nu > 0) \; .
\ee
For illustration, we may recall that for a uniform system
$$
\vp_0(\br) = \frac{1}{\sqrt{V}} = \sqrt{\frac{\rho}{N} } \; ,
\qquad |\vp_0(\br)|^2 = \frac{\rho}{N} \; ,
$$
hence inequality (22) is valid, with $\nu=1$. Another example is the
gas of atoms trapped in a potential $U(\br)\propto r^n$. Then [25],
inequality (22) is again valid, with $\nu=1+2/n$. Under condition (22),
one has
$$
\lim_{N\ra\infty} |\vp_0(\br) \vp^*_0(\br') | \leq
\lim_{N\ra\infty}\; \frac{const}{N^\nu} = 0 \; .
$$
Therefore, in thermodynamic limit, one can formally write down Eq.
(19) as
\be
\label{23}
\lim_{N\ra\infty} [\psi_0(\br),\;\psi_0^\dgr(\br')] = 0 \; ,
\ee
and the commutation relation (20) becomes of the Bose kind,
\be
\label{24}
\lim_{N\ra\infty} [\psi_1(\br),\;\psi_1^\dgr(\br')] =
\dlt(\br-\br') \; .
\ee
More rigorously, one should consider Eqs. (19) and (21) as equations
for distributions, which have sense being integrated with integrable
functions from the class of functions such that
$$
\lim_{N\ra\infty} \left | \int_V f(\br,\br')\; d\br d\br' \right |
< \infty \; .
$$
On this class of functions, taking into account condition (22), one
gets
$$
\lim_{N\ra\infty}  \int_V \; [\psi_0(\br),\;\psi_0^\dgr(\br')]
f(\br,\br') \; d\br d\br' = 0 \; ,
$$
$$
\lim_{N\ra\infty}  \int_V \; [\psi_1(\br),\;\psi_1^\dgr(\br')]
f(\br,\br') \; d\br d\br' = \lim_{N\ra\infty}
\int_V \dlt(\br-\br') f(\br,\br') \; d\br d\br'\; ,
$$
which explains Eqs. (23) and (24).

Thus, in thermodynamic limit, the condensate variable (15) can be treated as 
commuting and the operators of uncondensed particles (16) as usual Bose field 
operators. But for a finite system, one has to deal with the exact commutation 
relations (19), (20), and (21).

One says that BEC occurs in thermodynamic limit, when the number of condensed 
particles
\be
\label{25}
N_0 = \int <\psi_0^\dgr(\br)\psi_0(\br)> d\br \; = \;
<a_0^\dgr a_0>
\ee
satisfies the condensation criteria (8) or (12).

Substituting expansion (13), with notations (15), (16), and (17), into the 
density matrix (2), and assuming the quantum-number conservation condition
$$
< a_k^\dgr a_p> \; = \; \dlt_{kp} < a_k^\dgr a_k> \; ,
$$
we obtain the spectral resolution (9), in which $N_0$ is given by Eq. (25), 
and
$$
n_k \; = \; < a_k^\dgr a_k > \; .
$$

Nowhere throughout this section, gauge symmetry breaking has been either 
invoked or even mentioned. The spectral resolution (9) formally does not 
require the breaking of symmetry [26]. The BEC criteria (8) and (12) also 
are formulated without involving the notion of symmetry breaking. Because 
of this, one often states that BEC does not necessarily imply gauge symmetry 
breaking and that the overall phenomenon of BEC can perfectly be described 
without breaking the gauge symmetry. The latter conclusion, however, is not 
correct, since it does not follow from the fact that the spectral resolution
(9) and the BEC criteria (8) and (12) can be formulated not mentioning symmetry 
breaking. Strictly specking, Eqs. (9), (8), and (12) tell us nothing about 
whether gauge symmetry must be broken or not. To understand this, it is 
necessary to know the correct way of calculating the density matrix and all 
other averages.

\section{Methods of symmetry breaking}

For the practical purpose of describing BEC, one usually breaks gauge
symmetry. There are two ways of doing this, by the method of infinitesimal
sources and by the Bogolubov operator shift.

\subsection{Method of infinitesimal sources}

Let the system Hamiltonian $H[\psi]$, being a functional of the field operator 
$\psi$, be invariant under the gauge transformation $\psi\ra e^{i\al}\psi$, 
where $\al$ is a real number, so that
\be
\label{26}
H\left [ e^{i\al}\psi \right ] = H[\psi] \; .
\ee
To break the gauge symmetry, we can add to the Hamiltonian $H[\psi]$ a term 
explicitly breaking the symmetry, for instance, defining
\be
\label{27}
H_\ep[\psi] \equiv H[\psi] + \ep \; \sqrt{\rho} \; \int \left [
\psi_0^\dgr(\br) +\psi_0(\br) \right ] \; d\br \; .
\ee
Statistical averages for an equilibrium system are then averaged with the 
statistical operator
\be
\label{28}
\hat\rho_\ep \equiv
\frac{\exp\{-\bt H_\ep[\psi]\}}{{\rm Tr}\exp\{-\bt H_\ep[\psi]\}} \; ,
\ee
in which $\bt\equiv 1/T$ is inverse tepmerature. Thus, the statistical
average of an operator $\hat A$ is given by
\be
\label{29}
<\hat A >_\ep \equiv \; {\rm Tr}\hat\rho_\ep \hat A \; ,
\ee
where the trace is over the Fock space $\cF(\psi)$.

Since Hamiltonian (27) is not invariant under the gauge transformation,
the system symmetry is broken. The manifestation of the {\it broken
gauge symmetry} is
\be
\label{30}
< \psi(\br) >_\ep \; \neq 0 \; .
\ee

But it is important to find out what happens if the symmetry breaking
term is removed. If it is removed before thermodynamic limit, then
evidently
\be
\label{31}
\lim_{\ep\ra 0} <\psi(\br)>_\ep \; = 0 \; .
\ee
However, the average of $\psi$ may remain nonzero, if the symmetry breaking 
term is removed {\it after} thermodynamic limit. One says that {\it gauge 
symmetry is spontaneously broken}, when
\be
\label{32}
\lim_{\ep\ra 0}\; \lim_{N\ra\infty} < \psi(\br) >_\ep \; \neq 0 \; .
\ee
For a nonequilibrium system, the statistical operator (28) can be
treated as the initial form of the time-dependent statistical operator
$\hat\rho_\ep(t)$, so that $\hat\rho_\ep(0)=\hat\rho_\ep$.

The averages of type (32), defined through the double noncommuting limiting 
procedure $\lim_{\ep\ra 0}\lim_{N\ra\infty}<\ldots>$, are the Bogolubov 
quasiaverages [27,28]. It is also possible to define the thermodynamic 
quasiaverages [29--31] involving the sole limiting procedure, that of 
thermodynamic limit $\lim_{N\ra\infty}<\ldots>$. The latter procedure can 
also result in the spontaneous symmetry breaking, provided the infinitesimal 
source in thermodynamic limit tends to zero in the appropriate way [29--31].

\subsection{Bogolubov operator shift}

Another method of gauge symmetry breaking is based on the observation that 
in thermodynamic limit we come to the commutation relations (23) and (24). 
The first tells us that the operator $\psi_0(\br)$ is asymptotically equivalent 
to a nonoperator function, which can be expressed as the replacement
\be
\label{33}
\psi_0(\br) \; \ra \; \eta(\br) \; , \qquad
a_0 \; \ra \; \sqrt{N_0} \; ,
\ee
where $\eta(\br)$ is termed the condensate wave function. Respectively,
the field operator (17) is to be replaced as
\be
\label{34}
\psi(\br) \; \ra \; \hat\psi(\br) \equiv \eta(\br) +\psi_1(\br) \; ,
\ee
which is called the Bogolubov operator shift. The commutation relation
(24) signifies that the operator $\psi_1(\br)$ asymptotically acquires 
the usual Bose commutation relations. This agrees with the fact that if 
$\hat\psi(\br)$ is a Bose operator and $\eta(\br)$ is a nonoperator 
function, then $\psi_1(\br)$ in Eq. (34) has to be a Bose operator with 
the commutation relation
\be
\label{35}
[\psi_1(\br),\; \psi_1^\dgr(\br')] = \dlt(\br-\br') \; .
\ee

The orthogonality condition (18) transforms into
\be
\label{36}
\int \eta^*(\br)\psi_1(\br)\; d\br = 0 \; .
\ee
Thus, after the Bogolubov shift (34), instead of one field operator
$\psi(\br)$, as in Eq. (17), there appear two independent variables,
the condensate wave function $\eta(\br)$ and the field operator of
uncondensed particles $\psi_1(\br)$, satisfying the Bose commutation
relations (35).

The field operator $\psi_1^\dgr$ generates the Fock space
$\cF(\psi_1)$, which is orthogonal to the Fock space $\cF(\psi)$
generated by the operator $\psi^\dgr$. The Bogolubov shift (34)
realizes the unitary nonequivalent operator representations [32].
Consequently, any operator $\hat A[\psi]$, expressed through the
field operator $\psi$, is defined on the space $\cF(\psi)$, while
the corresponding operator $\hat A[\eta,\psi_1]$ acts on the space
$\cF(\psi_1)$.

Accomplishing the Bogolubov shift (34), we pass from the Hamiltonian
$H[\psi]$ to the Hamiltonian $H[\eta,\psi_1]$. Then the average
\be
\label{37}
<\hat A[\eta,\psi_1]>_\eta \; \equiv {\rm Tr}\; \hat\rho(\eta)
\hat A[\eta,\psi_1]
\ee
is defined as the trace over $\cF(\psi_1)$, with the statistical
operator
\be
\label{38}
\hat\rho(\eta) \equiv
\frac{\exp\{-\bt H[\eta,\psi_1]\}}
{{\rm Tr}\exp\{-\bt H[\eta,\psi_1]\}} \; .
\ee
Using average (37), one introduces the quantum-number conservation
condition
\be
\label{39}
< \psi_1(\br) >_\eta \; = 0 \; ,
\ee
which defines the operator average
\be
\label{40}
< \hat\psi(\br) >_\eta \; = \eta(\br)
\ee
as the order parameter of the system. The condensate wave function
is normalized to the number of condensed particles
\be
\label{41}
N_0 = \int |\eta(\br)|^2 \; d\br \; .
\ee
If the system is absolutely stable and the number $N_0$ satisfies
the condensation criteria (8) or (12), then there exists BEC.

The Hamiltonian (27), with the infinitesimal term, transforms to
$H_\ep[\eta,\psi_1]$. The related statistical operator becomes
\be
\label{42}
\hat\rho_\ep(\eta) \equiv
\frac{\exp\{-\bt H_\ep[\eta,\psi_1]\}}
{{\rm Tr}\exp\{-\bt H_\ep[\eta,\psi_1]\}} \; ,
\ee
where the trace is over $\cF(\psi_1)$. Similarly to Eq. (37), we now
have the average
\be
\label{43}
< \hat A[\eta,\psi_1]>_{\ep\eta}\; \equiv {\rm Tr} \; \hat\rho_\ep(\eta)
\hat A[\eta,\psi_1] \; .
\ee
Since
$$
\lim_{\ep\ra 0} H_\ep[\eta,\psi_1] =  H[\eta,\psi_1] \; ,
$$
averages (37) and (43) are connected through the relation
\be
\label{44}
\lim_{\ep\ra 0} <\hat A[\eta,\psi_1]>_{\ep\eta}\;  = \;
<\hat A[\eta,\psi_1]>_\eta \; .
\ee

The Bogolubov shift (34), because of condition (40), explicitly breaks gauge 
symmetry. When the infinitesimal term is added, then $<\psi_1(\br)>_{\ep\eta}$ 
is not zero. However, in compliance with Eq. (44), we have
$$
\lim_{\ep\ra 0} <\psi_1(\br)>_{\ep\eta} \; = \; <\psi_1(\br)>_\eta \;
= 0 \; .
$$
Therefore
\be
\label{45}
\lim_{\ep\ra 0} < \hat \psi(\br)>_{\ep\eta} \; =\; 
<\hat\psi(\br)>_\eta \; = \eta(\br) \; ,
\ee
which means that gauge symmetry is broken, provided $\eta(\br)$ is not zero. 
Thus, the Bogolubov shift (34) breaks the gauge symmetry without introducing 
infinitesimal sources.

\section{Sufficient and necessary conditions}

All observable and thermodynamic quantities of a statistical system can be 
expressed through some correlation functions. Let us consider the class of 
correlation functions
\be
\label{46}
C_\ep(\psi_0,\psi_1) \equiv \; < \ldots \psi_0^\dgr
\ldots \psi_1^\dgr \ldots \psi_0 \ldots \psi_1 \ldots >_\ep
\ee
consisting of the averages (29) of a product of the field operators
$\psi_0^\dgr$, $\psi_1^\dgr$, $\psi_0$, and $\psi_1$ in any order. And let 
us introduce the class of correlation functions
\be
\label{47}
C(\eta,\psi_1) \equiv \; < \ldots \eta^* \ldots \psi_1^\dgr
\ldots \eta \ldots \psi_1>_\eta
\ee
represented by averages (37). The order of the factors here is the same as 
in Eq. (46) with the change of $\psi_0$ by $\eta$. Also, we define the class 
of correlation functions
\be
\label{48}
C_\ep(\eta,\psi_1) \equiv \; <\ldots \eta^*\ldots \psi_1^\dgr
\ldots \eta \ldots \psi_1>_{\ep\eta}
\ee
given by the averages (43) of a product of $\eta^*$, $\psi_1^\dgr$, $\eta$, 
and $\psi_1$, whose order is the same as in Eq. (47).

The class of correlation functions (46) is defined for a system, where gauge 
symmetry is broken by means of the infinitesimal source. In this method, the 
limit $\ep\ra 0$ and the thermodynamic limit do not commute, so that
\be
\label{49}
[\lim_{\ep\ra 0}, \; \lim_{N\ra\infty}] C_\ep(\psi_0,\psi_1)
\neq 0 \; .
\ee
The class of correlation functions (48) is constructed by employing the 
Bogolubov shift (34). In the latter method, gauge symmetry is broken explicitly,
because of which the limits $\ep\ra 0$ and $N\ra\infty$ now commute,
\be
\label{50}
[\lim_{\ep\ra 0}, \; \lim_{N\ra\infty}] C_\ep(\eta,\psi_1) =
0 \; .
\ee

\subsection{Sufficient condition for condensation}

Bogolubov [28,33,34] showed that the methods of breaking gauge symmetry by 
means of infinitesimal sources and by the operator shift (34) are asymptotically
equivalent and the operator shift (34) is asymptotically exact in the following 
sense. Let us compare the classes defined in Eqs. (46), (47), and (48). The 
correlation functions from these classes will be called {\it similar} if they
have the same operator structure up to the change of $\psi_0$ by $\eta$.

\vskip 3mm

{\it Bogolubov theorem}.

\vskip 2mm

Similar correlation functions from the classes (46) and (48), in thermodynamic 
limit, coincide,
\be
\label{51}
\lim_{N\ra\infty} C_\ep(\psi_0,\psi_1) = \lim_{N\ra\infty}
C_\ep(\eta,\psi_1) \; .
\ee
This limiting equality holds for any $\ep$, including $\ep\ra 0$. Therefore, 
taking into account the commutation property (50), one has
\be
\label{52}
\lim_{\ep\ra 0}\; \lim_{N\ra\infty} C_\ep(\psi_0,\psi_1) =
\lim_{N\ra\infty} C(\eta,\psi_1) \; .
\ee
The limits in the left-hand side of Eq. (52) do not commute.

From the Bogolubov theorem, it follows, in particular, that
\be
\label{53}
\lim_{N\ra\infty}\left | < \psi_0(\br)>_\ep \right |^2 =
\lim_{N\ra\infty} < \psi_0^\dgr(\br) \psi_0(\br)>_\ep \; .
\ee
This is evident, since Eq. (53), with property (51), reduces to the trivial 
identity $|\eta|^2=|\eta|^2$. For a uniform system, Eq. (53) becomes
\be
\label{54}
\lim_{V\ra\infty} \; \frac{1}{V}\; |<a_0>_\ep|^2 =
\lim_{V\ra\infty} \; \frac{1}{V} < a_0^\dgr a_0 >_\ep \; .
\ee
This again immediately follows from the Bogolubov theorem (51) and, according 
to Eq. (33), reduces to the identity $\rho_0=\rho_0$. Equality (54) has also 
been obtained by Belyaev [35] and Ginibre [36] and recently an elegant proof 
was given by Lieb et al. [37--39].

Equalities (53) and (54) are valid for any $\ep$, including $\ep\ra 0$.
Defining the local condensate density
\be
\label{55}
\rho_0(\br) \equiv \lim_{\ep\ra 0} \; \lim_{N\ra\infty}
< \psi_0^\dgr(\br) \psi_0(\br) >_\ep \; ,
\ee
we find from Eq. (53)
\be
\label{56}
\lim_{\ep\ra 0} \; \lim_{N\ra\infty} | < \psi_0(\br)>_\ep |^2
= \rho_0(\br) \; .
\ee
The latter equation shows us that if gauge symmetry is locally spontaneously 
broken, so that the left-hand side of Eq. (56) is not identically zero, then 
there exists local BEC in the sense that $\rho_0(\br)>0$ at least for some 
$\br$. And if there is no spontaneous symmetry breaking, at least locally, so 
that the left-hand side of Eq. (56) is identically zero, then there is no BEC, 
since $\rho_0(\br)=0$. When spontaneous breaking of gauge symmetry occurs 
globally, in the sense that
\be
\label{57}
\lim_{\ep\ra 0} \; \lim_{V\ra\infty} \; \frac{1}{V} \;
\int_V \; | < \psi_0(\br) >_\ep |^2 \; d\br \; > \; 0 \; ,
\ee
then there exists BEC in the usual sense of criteria (8) or (12), since then
$$
\lim_{V\ra\infty}\; \frac{1}{V} \; \int_V \; \rho_0(\br)\;
d\br \; > 0\; .
$$
For a uniform system, as is clear, the local and global BEC conditions are 
equivalent.

Note that breaking of gauge symmetry can interchangeably be considered for 
the averages of either $\psi$ or $\psi_0$, since $\psi=\psi_0+\psi_1$ and 
from the Bogolubov theorem we have
$$
\lim_{\ep\ra 0} \; \lim_{N\ra\infty} < \psi_1(\br) >_\ep \; =
\lim_{\ep\ra 0} \; \lim_{N\ra\infty} < \psi_1(\br) >_{\ep\eta}\; =
\lim_{N\ra\infty} < \psi_1(\br) >_\eta \; = 0 \; ,
$$
where condition (39) is also taken into account. Therefore
\be
\label{58}
\lim_{\ep\ra 0} \; \lim_{N\ra\infty} < \psi(\br) >_\ep \; =
\lim_{\ep\ra 0} \; \lim_{N\ra\infty} < \psi_0(\br) >_\ep \; .
\ee

The asymptotic exactness of the Bogolubov shift (34) has also been proved 
for thermodynamic potentials by Ginibre [36] and Lieb et al. [37--39]. Let 
us consider the thermodynamic potentials
\be
\label{59}
\Om_\ep \equiv - T \ln \;
{\rm Tr}\exp \{ -\bt H_\ep[\psi] \} \; ,
\ee
with the trace over $\cF(\psi)$, and
\be
\label{60}
\Om_\ep(\eta) \equiv - T \ln \;
{\rm Tr}\exp \{ -\bt H_\ep[\eta,\psi_1] \} \; ,
\ee
with the trace over $\cF(\psi_1)$. And let $\eta$ be the minimizer for the 
potential
$$
\Om_\ep(\eta) = {\rm inf}_x \; \Om_\ep(x) \; .
$$
Then, for sufficiently general conditions on the interaction potential (see 
details in Refs. [36--39]), one has the {\it Ginibre theorem}
\be
\label{61}
\lim_{V\ra\infty}\; \frac{1}{V} \left [ \Om_\ep(\eta) -
\Om_\ep \right ] = 0 \; .
\ee
Ginibre [36] proved this theorem for $\ep>0$ as well as for $\ep=0$.

The Bogolubov and Ginibre theorems show that two ways of breaking gauge 
symmetry, by means of infinitesimal sources and by the Bogolubov shift 
(34), are asymptotically equivalent and that the Bogolubov shift (34) is 
asymptotically exact in the sense of Eqs. (51), (52), and (61). It also 
follows that the {\it spontaneous gauge symmetry breaking is a sufficient 
condition for the existence of BEC}.

\subsection{Necessary condition for condensation}

The fundamental question is whether the gauge symmetry breaking would also 
be the necessary condition for BEC. This problem has been discussed in a 
number of works, of which we mention here just a few [1,2,22,26,40]. One 
often states that gauge symmetry breaking is not necessary for BEC and that 
the letter can occur without any symmetry breaking. This, however, is a 
delusion, though a widespread one.

Suppose, we do not break gauge symmetry and work with the statistical operator
\be
\label{62}
\hat\rho \equiv
\frac{\exp\{-\bt H[\psi]\}}{{\rm Tr}\exp\{ -\bt H[\psi]\} } \; ,
\ee
in which $H[\psi]$ is a gauge-symmetric Hamiltonian. The related gauge-symmetric
averages are defined as
\be
\label{63}
< \hat A > \; \equiv \; {\rm Tr}\hat\rho \hat A \; .
\ee
For a uniform Bose-condensed system, Roepstorff [41] proved the inequality
\be
\label{64}
\lim_{N\ra\infty}\; \frac{<a_0^\dgr a_0>}{N} \;
\leq \; \lim_{\ep\ra 0} \; \lim_{N\ra\infty} \;
\frac{|<a_0>_\ep|^2}{N} \; .
\ee
In his proof, he used the assumption that, in thermodynamic limit, $H[\psi]$ 
commutes with $\psi_0$. This is a plausible assumption, since, as is discussed 
above, in thermodynamic limit, $\psi_0$ can be replaced by a nonoperator 
function $\eta$. A more general proof of inequality (64) was given by Lieb 
et al. in a series of papers [37--39]. Actually, the main requirement, used 
by Lieb et al. [37--39], was the existence of the Fourier transform for the
interaction potential. The case of hard-core potentials can also be taken care 
of by cutting off the potential at some finite value that is taken to infinity 
at the end of the calculations [37--39].

In the coordinate representation, the Roepstorff inequality (64) can be written 
as
\be
\label{65}
\lim_{V\ra\infty}\; \frac{1}{V} \; \int_V
< \psi_0^\dgr(\br) \psi_0(\br) > d\br \; \leq \;
\lim_{\ep\ra 0} \; \lim_{V\ra\infty} \;
\frac{1}{V} \; \int_V \; | < \psi_0(\br)>_\ep |^2 d\br \; .
\ee
When the left-hand sides of Eqs. (64) or (65) are nonzero, this means the 
existence of BEC, in compliance with the BEC criteria (8) or (12). But then 
the right-hand sides of Eqs. (64) and (65) are also nonzero, which implies 
spontaneous symmetry breaking. That is, the {\it existence of BEC necessarily 
assumes spontaneous gauge symmetry breaking}.

Equalities (64) and (65) are the direct analogs of the Griffiths [42] 
inequality
$$
\lim_{N\ra\infty} <\hat\bS^2 > \;\; \leq \;
\lim_{\ep\ra 0} \; \lim_{N\ra\infty} | < \hat\bS >_\ep |^2
$$
for magnetic systems, where
$$
\hat\bS \equiv \frac{1}{N} \; \sum_{i=1}^N \hat\bS_i
$$
is the mean spin operator and the symmetry breaking term is $\ep N\hat\bS^z$, 
so that $\ep$ plays the role of an external magnetic field. The Griffiths 
inequality tells us that the appearance of magnetization necessarily implies 
the spontaneously broken spin-rotational symmetry $O(3)$.

In this way, from the Bogolubov theorem, formulated in Eq. (51), equalities 
(53) to (56) follow, which have also been proved by Ginibre [36] and by Lieb 
et al. [37--39]. These equalities show that the spontaneous gauge symmetry 
breaking is a {\it sufficient} condition for the occurrence of BEC. And 
inequalities (64) and (65), proved by Roepstorff [41] and by Lieb et al. 
[37--39], establish that the existence of BEC is {\it necessarily} 
accompanied by the spontaneous gauge symmetry breaking. Thence, we come to 
the conclusion:

\vskip 2mm

{\it The spontaneous gauge symmetry breaking is the necessary and sufficient 
condition for the existence of Bose-Einstein condensate}.

\subsection{Illustration by mean-field model}

In order to illustrate the above conclusion explicitly, let us consider a 
simple mean-field type model with the Hamiltonian
\be
\label{66}
H = \sum_k (\om_k - \mu) a_k^\dgr a_k \; ,
\ee
where $a_k$ and $a_k^\dgr$ are Bose operators and $\om_k$ is an effective 
spectrum, such that $\om_k-\mu\geq 0$ for all multi-indices $k$. For a 
uniform system, the index $k$ becomes the momentum $\bk$. But, in general, 
the system can be of any nature.

For temperatures $T$ above the condensation temperature $T_c$, there is no 
need to break gauge symmetry. Then the particle distribution is
\be
\label{67}
< a_k^\dgr a_k > \; = \; \left [ e^{\bt(\om_k-\mu)} - 1
\right ]^{-1}
\ee
for any $k$. The appearance of a Bose condensate presupposes that the 
ground-state level $\om_0$ gets macroscopically populated, as a result of 
which the number of particles on this level,
\be
\label{68}
< a_0^\dgr a_0 > \; = \; \left [ e^{\bt(\om_0-\mu)} - 1
\right ]^{-1} \; ,
\ee
becomes proportional to the total number of particles $N$. In order that 
$<a_0^\dgr a_0>\propto N$, it is necessary that
\be
\label{69}
\lim_{N\ra\infty} (\om_0 - \mu) \; \ra \; +0 \qquad
(T \leq T_c) \; .
\ee
Hence, Eq. (69) is a necessary condition for BEC.

Breaking gauge symmetry by the infinitesimal term, as in Eq. (27), we now have
\be
\label{70}
H_\ep \equiv H + \ep \sqrt{N} \left ( a_0^\dgr + a_0 \right ) \; .
\ee
Hamiltonian (70), with $H$ from Eq. (66), can be diagonalized by means of 
the canonical transformation
\be
\label{71}
a_0 = b_0 + \frac{\ep\sqrt{N}}{\mu-\om_0} \; .
\ee
Then Eq. (70) transforms into
\be
\label{72}
H_\ep = (\om_0 - \mu) b_0^\dgr b_0 + \sum_{k\neq 0}
( \om_k - \mu ) a_k^\dgr a_k + \frac{\ep^2N}{\mu-\om_0} \; .
\ee
The latter Hamiltonian is gauge-invariant with respect to the change
$b_0\ra b_0 e^{i\al}$ and $a_k\ra a_ke^{i\al}$. Therefore,
\be
\label{73}
< b_0 >_\ep \; = 0 \; ,
\ee
where the averaging is with Hamiltonian (72) and, similarly,
\be
\label{74}
< a_k >_\ep \; = 0 \qquad (k\neq 0) \; .
\ee
From Eq. (71), we find
\be
\label{75}
< a_0 >_\ep \; = \frac{\ep\sqrt{N}}{\mu-\om_0} \; .
\ee
For the bilinear averages, we get
\be
\label{76}
< b_0^\dgr b_0>_\ep \; = \; < a_0^\dgr a_0> \; ,
\ee
whose form is given by Eq. (68), and
\be
\label{77}
< a_k^\dgr a_k>_\ep \; = \; < a_k^\dgr a_k> \qquad (k \neq 0) \; ,
\ee
defined by expression (67).

If there is no BEC, hence, condition (69) does not hold, then there is no 
spontaneous gauge symmetry breaking, since
\be
\label{78}
\lim_{\ep\ra 0}\; \lim_{N\ra\infty}\;
\frac{<a_0>_\ep}{\sqrt{N}} = 0 \qquad (\mu\neq \om_0) \; .
\ee
The number of condensed particles is, generally, different being calculated 
using the averages with gauge symmetry breaking or without it. Let us denote 
the former as
\be
\label{79}
N_0(\ep) \equiv \; < a_0^\dgr a_0>_\ep \qquad (\ep\neq 0)
\ee
and the latter as
\be
\label{80}
N_0(0) \equiv \; < a_0^\dgr a_0> \qquad (\ep\equiv 0) \; .
\ee
In agreement with transformation (71), we have
\be
\label{81}
N_0(\ep) = \; < b_0^\dgr b_0>_\ep + \frac{\ep^2N}{(\mu-\om_0)^2} \; .
\ee
From here, keeping in mind equality (76), we get
\be
\label{82}
< a_0^\dgr a_0> \; = N_0(0) \; \leq \; N_0(\ep) = \;
< a_0^\dgr a_0>_\ep \; .
\ee

Let below $T_c$ the condensate arise, so that $N_0(\ep)>0$ and
condition (69) holds. Then, for large $N$, from Eq. (81), we obtain
\be
\label{83}
\mu \simeq \om_0 \; - \; \frac{T}{2N_0(\ep)} \left [ 1 +
\sqrt{1+(2\bt\ep)^2N_0(\ep)N} \right ] \; .
\ee
For $\ep\neq 0$ and $N\ra\infty$, Eq. (83) simplifies to
\be
\label{84}
\mu \simeq \om_0 - \ep\; \sqrt{ \frac{N}{N_0(\ep)}} \; .
\ee
Using this in transformation (71), we have
\be
\label{85}
a_0 \simeq b_0 -\sqrt{N_0(\ep)} \; .
\ee
Then, in view of Eq. (73), we obtain
\be
\label{86}
<a_0>_\ep \; \simeq -\sqrt{N_0(\ep)} \; .
\ee
Equation (85) is analogous to the Bogolubov shift, and Eq. (86) yields
\be
\label{87}
\lim_{N\ra\infty} \; \frac{|<a_0>_\ep|^2}{N} =
\lim_{N\ra\infty}\; \frac{N_0(\ep)}{N} \; .
\ee
The latter equality, together with definition (79), takes the form
\be
\label{88}
\lim_{N\ra\infty}\; \frac{|<a_0>_\ep|^2}{N} =
\lim_{N\ra\infty} \; \frac{<a_0^\dgr a_0>_\ep}{N} \; ,
\ee
which is equivalent to the Ginibre-Lieb equality (54). Equation (88) tells 
us that asymptotically
\be
\label{89}
< a_0^\dgr a_0>_\ep \; \simeq |<a_0>_\ep|^2 \qquad (N\ra\infty) \; .
\ee
Taking in Eq. (88) the limit $\ep\ra 0$, we see that the spontaneous gauge 
symmetry breaking implies the existence of BEC, being a {\it sufficient} 
condition for the latter.

On the other hand, from Eqs. (82) and (88), with definitions (79) and (80), 
we obtain
\be
\label{90}
\lim_{N\ra\infty} \; \frac{<a_0^\dgr a_0>}{N} \; \leq \;
\lim_{N\ra\infty}\; \frac{|<a_0>_\ep|^2}{N} \; .
\ee
For $\ep\ra 0$, Eq. (90) becomes the Roepstorff-Lieb inequality (64), which 
establishes the {\it necessity} of spontaneous gauge symmetry breaking for 
the BEC existence.

\section{Conservation of the number of particles}

If the spontaneous gauge symmetry breaking is a necessary condition for BEC, 
then where is a flaw in the standard argument of those who claim that BEC 
does not need gauge symmetry breaking and can be realized without the latter? 
This argument is as follows. When gauge symmetry has been broken, the 
number-of-particle operator $\hat N$ does not commute with the Hamiltonian 
$H_\ep$, hence, $\hat N$ is not an integral of motion. However, in experiments, 
for instance with trapped atoms, the number of particles $N$ can be well 
controlled and kept constant during all the process of measurement.

\subsection{Conservation in the sense of quasiaverages}

The flaw in this argument is that it confuses the {\it operator} $\hat N$ with 
the observable quantity, which is the {\it average} $N=<\hat N>$. There is no 
any contradiction between the noncommutativity of an operator with a Hamiltonian,
in which a symmetry is broken by {\it infinitesimal} sources, and the fact that 
the average of this operator can be well controlled and kept constant in 
experiments. The situation for BEC is analogous to that for magnetic systems. 
In the latter, the appearance of magnetization is accompanied by the spontaneous
breaking of spin-rotational symmetry. When the latter is broken, the spin
operator $\hat\bS$ does not commute with the Hamiltonian. This, however, 
in no way forbids the magnetization, proportional to the spin average 
$<\hat\bS>$, from being well controlled, accurately measured, and kept 
constant in experiments.

To demonstrate what is said above mathematically, let us consider the 
number-of-particle operator
\be
\label{91}
\hat N \equiv \int \psi^\dgr(\br) \psi(\br) \; d\br \; .
\ee
This operator commutes with the gauge-symmetric Hamiltonian $H[\psi]$,
\be
\label{92}
[H[\psi],\; \hat N] = 0\; .
\ee
But after breaking the gauge symmetry with the infinitesimal source, the 
Hamiltonian $H_\ep[\psi]$ in Eq. (27) does not commute with $\hat N$. Keeping 
in mind the separation of the field operator into two parts, as in Eq. (17), 
we have
$$
[\psi_0(\br),\; \hat n(\br')] = \vp_0(\br) \vp_0^*(\br') \psi(\br') \; ,
$$
where $\hat n(\br)\equiv \psi^\dgr(\br)\psi(\br)$. From here,
$$
[\psi_0(\br),\; \hat N] = \psi_0(\br) \; .
$$
Therefore the commutator of $H_\ep\equiv H_\ep[\psi]$ with operator (91) is
\be
\label{93}
[H_\ep,\;\hat N] =\ep\sqrt{\rho} \int \left [ \psi_0^\dgr(\br) +
\psi_0(\br) \right ]\; d\br \; .
\ee
The number-of-particle operator (91) is now not an integral of motion, 
since the right-hand side of the Heisenberg equation
\be
\label{94}
i\; \frac{d}{dt} \; \hat N = [ \hat N,\; H_\ep]
\ee
is not zero, being given by Eq. (93). But the observable quantities are the 
averages of operators. Hence, we should consider the averaged Eq. (94), which 
gives
\be
\label{95}
\frac{d}{dt} <\hat N>_\ep \; =  i< [H_\ep,\; \hat N]>_\ep \; .
\ee
In view of Eq. (93), we get
\be
\label{96}
\frac{d}{dt} <\hat N>_\ep \; = i\ep\sqrt{\rho} \; \int <\psi_0^\dgr(\br)
- \psi_0(\br)>_\ep\; d\br \; .
\ee
To correctly define thermodynamic limit, we have to work with reduced
quantities. For example, Eq. (95) can be rewritten as
\be
\label{97}
\lim_{V\ra\infty}\; \frac{d}{dt}\; \frac{<\hat N>_\ep}{V} = i\;
\lim_{V\ra\infty}\; \frac{<[H_\ep,\hat N]>_\ep}{V} \; .
\ee
Using Eq. (93) and the Bogolubov theorem (51), we find
\be
\label{98}
\lim_{V\ra\infty}\; \frac{d}{dt}\; \frac{<\hat N>_\ep}{V} = i\;
\ep\; \lim_{V\ra\infty}\; \frac{\sqrt{\rho}}{V} \;
\int_V <\eta^*(\br) - \eta(\br)>_\ep \; d\br \; .
\ee
Breaking gauge symmetry by infinitesimal sources requires to define all 
observable quantities as Bogolubov quasiaverages. For the condensate function 
$\eta(\br)$, rememebring normalization (41) and using the Cauchy-Schwarz 
inequality, we have
$$
\left | \int \eta(\br)\; d\br \right |^2 \; \leq \; N_0 V \; .
$$
Because of this,
$$
\left | \frac{\sqrt{\rho}}{V} \; \int_V <\eta^*(\br) -
\eta(\br)>_\ep \; d\br \right | \; \leq 2 \sqrt{\rho\rho_0} \; .
$$
Therefore, from Eq. (98), we obtain
\be
\label{99}
\lim_{\ep\ra 0} \; \lim_{V\ra\infty} \;
\frac{d}{dt}\; \frac{<\hat N>_\ep}{V} = 0 \; .
\ee
This means that the average number of particles $N$, defined by the relation
\be
\label{100}
\lim_{V\ra\infty}\; \frac{N}{V} = \lim_{\ep\ra 0} \;
\lim_{V\ra\infty}\; \frac{<\hat N>_\ep}{V} \; ,
\ee
does not change in time and, being fixed at the initial time, can be kept 
constant during any experiment. The number of particles $N$, according to 
Eq. (100), can be thought as being defined by the asymptotic form 
$N\simeq<\hat N>_\ep$, assuming that $\ep\ra 0$ and $N\ra\infty$. Thus, 
though the operator $\hat N$ does not commute with $H_\ep$, the observable 
number of particles $N$ is a well controlled quantity that can be fixed with 
any desired accuracy.

The form of Eq. (98) hints that the influence of the gauge-symmetry breaking 
source is in the appearance of a phase in $<\hat N>_\ep$, proportional to $\ep$.
This is easy to illustrate for a uniform system, when $\eta(\br)=\sqrt{\rho_0}$.
Then asymptotically, as $N\ra\infty$, we have
\be
\label{101}
< [ H_\ep,\; \hat N] >_\ep \; \sim 2\ep\sqrt{n_0}\; N \; ,
\ee
where
\be
\label{102}
n_0 \equiv \lim_{N\ra\infty}\; \frac{N_0}{N}
\ee
is the condensate fraction. Keeping in mind that, in accordance with Eq. (100), 
asymptotically $N\simeq<\hat N>_\ep$, we obtain from Eq. (98) the asymptotic 
equation
\be
\label{103}
\frac{dN}{dt} \sim 2\ep\sqrt{n_0} \; N \; .
\ee
From here
\be
\label{104}
N(t) \sim N(0) \exp\left ( 2\ep\sqrt{n_0}\; t\right ) \; .
\ee
Passing again to quasiaverages, we get
\be
\label{105}
\lim_{\ep\ra 0} \; \lim_{V\ra\infty}\; \frac{N(t)}{N(0)}
= 1 \; .
\ee
This confirms once more that the observable number of particles can always 
be treated as a well defined, conserved, and precisely controlled quantity.

\subsection{Conservation under Bogolubov shift}

Moreover, the number-of-particle operator can be explicitly an integral of 
motion, at the same time with gauge symmetry being broken. This can be 
realized in the representation employing the Bogolubov shift. The initial 
Hamiltonian $H[\psi]$, which is a functional of the field operators $\psi$ 
and $\psi^\dgr$, is gauge invariant, according to property (26). Accomplishing 
the Bogolubov shift (34) leaves the operator structure of the new 
Hamiltonian $H[\hat\psi]\equiv H[\eta,\psi_1]$ the same as that of 
$H[\psi]$. Hence, the Hamiltonian $H[\eta,\psi_1]$ is invariant under 
the transformation $\hat\psi\ra\hat\psi e^{i\al}$, that is, 
$\eta\ra\eta e^{i\al}$ and $\psi_1\ra\psi_1 e^{i\al}$. As far as 
$H[\psi]$, by assumption (26), is gauge invariant, it commutes with 
the number-of-particle operator $\hat N[\psi]$. Since the Bogolubov 
shift (34) does not change the operator structure, except for replacing 
$\psi$ by $\hat\psi$, the Hamiltonian $H[\eta,\psi_1]$ commutes with the 
number-of-particle operator
\be
\label{106}
\hat N[\eta,\psi_1] \equiv \int 
\hat\psi^\dgr(\br) \hat\psi(\br)\; d\br \; ,
\ee
so that
\be
\label{107}
\left [ \hat N[\eta,\psi_1], \; H[\eta,\psi_1]
\right ] = 0 \; .
\ee
Operator (106) can be written as the sum
\be
\label{108}
\hat N[\eta,\psi_1] = N_0 + \hat N_1 
\ee
of the number of condensed particles (41) and the operator
\be
\label{109}
\hat N_1 \equiv \int \psi_1^\dgr(\br) \psi_1(\br)\; d\br
\ee
for the number of uncondensed particles. Thus the number-of-particle operator 
(106), in view of commutator (107), is the integral of motion, which means that 
the number of particles
\be
\label{110}
N \equiv \; < \hat N[\eta,\psi_1]>_\eta \; ,
\ee
with the average defined in Eq. (37), is a conserved quantity, which can be 
measured and fixed with the desired accuracy. At the same time, the gauge 
symmetry, in compliance with Eq. (40), is broken.

In order to understand how it may happen that the number-of-particle operator 
is the integral of motion, but the gauge symmetry is broken, we should remember 
that symmetry can be broken not solely in the Hamiltonian, by adding 
infinitesimal sources, but also by constructing an operator representation on 
a space of microstates with broken symmetry [31]. To be more precise, we can 
invoke the decomposition theory [43--45].

For a finite system, the space of microstates is the Fock space $\cF(\psi)$ 
generated by the field operator $\psi^\dgr$. The space is gauge invariant, 
in the sense that $\cF\left (\psi e^{i\al}\right)=\cF(\psi)$. The Fock space 
$\cF(\psi_1)$, generated by $\psi_1^\dgr$, is not gauge invariant. Actually, 
one can define an infinite number of spaces
\be
\label{111}
\cF_\al \equiv \cF\left (\psi_1 e^{i\al}\right ) \qquad 
(0 \leq \al < 2\pi) \; .
\ee
In thermodynamic limit, the sole space $\cF(\psi)$ disintegrates into the 
direct integral
\be
\label{112}
\cF = \int^\oplus \; \cF_a \; \frac{d\al}{2\pi}
\ee
of the partial spaces (111).

On each of subspaces (111), one can define a representation of the operator 
algebra, whose members are denoted as
\be
\label{113}
\hat A_\al \equiv \hat A
\left [ \eta e^{i\al}, \; \psi_1 e^{i\al} \right ] \; .
\ee
For this representation, we can construct the corresponding statistical 
averages by introducing the statistical operator
\be
\label{114}
\hat\rho_\al(\eta) \equiv 
\frac{\exp(-\bt H_\al)}{{\rm Tr}_{\cF_\al}\exp(-\bt H_\al)} \; ,
\ee
in which $H_\al\equiv H\left[\eta e^{i\al},\psi_1 e^{i\al}\right]$, 
according to Eq. (113). Then we have
\be
\label{115}
<\hat A_\al >_{\eta\al} \; \equiv 
{\rm Tr}_{\cF_\al} \hat\rho_\al(\eta) \hat A_\al \; .
\ee
It is easy to notice that
$$
\lim_{\al\ra 0} <\hat A_\al>_{\eta\al} \; = \; 
<\hat A[\eta,\psi_1]>_\eta \; ,
$$
with the right-hand side given by Eq. (37). The average of an operator 
$\hat A$, defined on the gauge-invariant space (112), is decomposed into 
the integral
\be
\label{116}
< \hat A >_\cF \; = \int_0^{2\pi}  < \hat A_\al>_{\eta\al} \;
\frac{d\al}{2\pi}
\ee
over gauge-noninvariant averages defined for subspaces (111).

In conformity with condition (39), we require that
\be
\label{117}
< \psi_1(\br)>_{\eta\al} \; = 0 \; .
\ee
Then the gauge-invariant average of $\hat\psi$ is
\be
\label{118}
< \hat\psi(\br)>_\cF \; \equiv \int_0^{2\pi} \; 
\eta(\br) e^{i\al} \; \frac{d\al}{2\pi} = 0 \; ,
\ee
as it should be. However, the average of the representation $\hat\psi_\al(\br)
\equiv\hat\psi(\br)e^{i\al}$ over the gauge-noninvariant space (111) is not 
zero,
\be
\label{119}
< \hat\psi_\al(\br) >_{\eta\al} \; = \eta(\br) e^{i\al} \; ,
\ee
which implies gauge symmetry breaking. This is the way how the gauge symmetry 
of a system can be broken, despite that the number-of-particle operator remains 
the integral of motion.

One may also notice that, under the assumption that $\cF(\psi)$ tends, in 
thermodynamic limit, to $\cF$, given by decomposition (112), then
\be
\label{120}
\lim_{N\ra\infty} <\psi_0^\dgr(\br) \psi_0(\br)> \; = |\eta(\br)|^2 \; .
\ee
This again proves that BEC is equivalent to gauge symmetry breaking. In other 
words, the latter is a necessary condition for the existence of BEC.

\section{Negligible condensate fluctuations}

In recent years much attention has been given to the study of condensate 
fluctuations (see review article [5]). One often claims that these fluctuations 
are anomalous or even catastrophic. However, if to calculate these fluctuations 
correctly, not forgetting about the broken gauge symmetry, which is a necessary 
condition, then they are neither anomalous nor catastrophic, but, moreover,
they are negligible in thermodynamic limit.

One describes the condensate fluctuations as the fluctuations associated with 
the condensate number-of-particle operator
\be
\label{121}
\hat N_0 \equiv \int \psi_0^\dgr(\br) \psi_0(\br)\; d\br \; .
\ee
The fluctuations are characterized by the dispersion of operator (121),
\be
\label{122}
\Dlt^2_\ep(\hat N_0) \equiv \; <\hat N_0^2>_\ep -
<\hat N_0>_\ep^2 \; .
\ee
The dispersion of an extensive operator is not, of course, an observable 
quantity by its own. The observable quantity is the ratio
\be
\label{123}
\lim_{N\ra\infty} \; \frac{\Dlt^2(\hat N_0)}{N} \equiv
\lim_{\ep\ra 0}\; \lim_{N\ra\infty}\; \frac{\Dlt^2_\ep(\hat N_0)}{N} \; ,
\ee
which is analogous to the definition of other extensive observables [46,47]. 
Denoting
\be
\label{124}
\hat n_0(\br) \equiv \psi_0^\dgr(\br) \psi_0(\br) \; ,
\ee
we can write down dispersion (122) as
\be
\label{125}
\Dlt^2_\ep(\hat N_0) = \int \left [ <\hat n_0(\br) \hat n_0(\br')>_\ep 
- <\hat n_0(\br)>_\ep <\hat n_0(\br')>_\ep \right ] \; d\br d\br' \; .
\ee
In thermodynamic limit, according to the Bogolubov theorem, we should make the 
replacement $\psi_0(\br)\ra\eta(\br)$, hence
$$
\hat n_0(\br) \; \ra \; \rho_0(\br) \equiv |\eta(\br)|^2 \; .
$$
Then Eq. (125) immediately gives $\Dlt^2_\ep(\hat N_0)\simeq 0$.

One may ask what happens if one represents dispersion (125) is a slightly 
different form following from the commutation relation (19)? Using the latter, 
we can write
\be
\label{126}
\hat n_0(\br) \hat n_0(\br') = \psi_0^\dgr(\br) \psi_0^\dgr(\br') 
\psi_0(\br')\psi_0(\br) + 
\psi_0^\dgr(\br)\psi_0(\br')\vp_0(\br)\vp_0^*(\br') \; .
\ee
We should expect that the result for dispersion (125) could not depend on the 
used commutation relation (19), since the condensate operators $\psi_0(\br)$ 
and $\psi_0^\dgr(\br')$, in agreement with Eq. (23), commute in thermodynamic 
limit. More accurately, there is the property
\be
\label{127}
\lim_{N\ra\infty} < \left [ \psi_0(\br),\psi_0^\dgr(\br')
\right ] >_\ep = 0\; .
\ee

One should exercise extreme caution integrating expressions containing the 
commutation relations (19). As is explained in Section 2, such expressions 
have to be understood in the sense of equations for distributions. Their 
integration is correctly defined only for the integration with an integrable 
function, as is specified in Section 2. Thus the integration of Eq. (127) 
over the spatial variables has sense only for the case
\be
\label{128}
\lim_{N\ra\infty} \int_V < \left [ 
\psi_0(\br),\psi_0^\dgr(\br')\right ]>_\ep
f(\br,\br')\; d\br \; d\br' = 0 \; ,
\ee
where $f(\br,\br')$ is an integrable function, such as defined in subsection 
2.2. The integration without such a function can yield wrong results. For 
example, we could get
$$
\int_V < \left [ \psi_0(\br),\psi_0^\dgr(\br') \right ] >_\ep 
d\br \; \ra \; 1
$$
for large $N\ra\infty$, which would contradict the commutativity property 
(23). To exclude the appearance of such incorrect terms, one has to employ the 
regularized limit
$$
\lim_{N\ra\infty}\; \frac{\Dlt^2_\ep(\hat N)}{N} =
$$
\be
\label{129}
= \lim_{f\ra 1}\; \lim_{N\ra\infty}\; \frac{1}{N} \; \int_V
\left [ <\hat n_0(\br)\hat n_0(\br')>_\ep - <\hat n_0(\br)>_\ep 
<\hat n_0(\br')>_\ep \right ] f(\br,\br')\; d\br \; d\br' \; .
\ee
Reducing the regularizing function $f(\br,\br')$ to unity can be done only 
{\it after} thermodynamic limit, but not before, since
$$
\left [ \lim_{f\ra 1}, \; \lim_{N\ra\infty}\right ] \neq 0 \; .
$$

Suppose, we decide to use form (126) in dispersion (125). Employing the 
equality $\psi_0(\br)=a_0\vp_0(\br)$, the Bogolubov replacement 
$a_0\ra\sqrt{N_0}$, condition (22), and taking care of correct definitions, 
we have
$$
\lim_{N\ra\infty} \; \frac{1}{N} \left | \int_V 
< \psi_0^\dgr(\br)\psi_0(\br')>_\ep \vp_0(\br)\vp_0^*(\br') 
f(\br,\br')\; d\br\; d\br'\right | \; \leq
$$
$$
\leq \; \lim_{N\ra\infty} \left | \rho_0 \; \frac{const}{N^{2\nu}} \; 
\int_V \; f(\br,\br')\; d\br\; d\br' \right | = 0 \; .
$$
Thence, the additional term in Eq. (126), caused by the commutation relations 
(19), does not contribute to the reduced dispersion (129). In any case, we have
\be
\label{130}
\lim_{N\ra\infty} \; \frac{\Dlt^2_\ep(\hat N_0)}{N} = 0 \; .
\ee
This means that the condensate fluctuations are negligible in thermodynamic 
limit.

If we would employ representation (15) in the definition of operator (121), we 
would get
\be
\label{131}
\hat N_0 = \int \hat n_0(\br)\; d\br = a_0^\dgr a_0 \; .
\ee
Hence,
\be
\label{132}
\hat N_0^2 = a_0^\dgr a_0 a_0^\dgr a_0 \; .
\ee
Using here the commutation relation $[a_0,a_0^\dgr]=1$, we have another form
\be
\label{133}
\hat N_0^2 = a_0^\dgr a_0^\dgr a_0 a_0 + a_0^\dgr a_0 \; .
\ee
It looks like Eqs. (132) and (133) give different answers for dispersion (125). 
For instance, keeping in mind the Bogolubov replacement $a_0\ra\sqrt{N_0}$ and 
using form (132), we have $\Dlt^2_\ep(\hat N_0)=0$, in agreement with Eq. (130).
At the same time, using form (133), we would get $\Dlt^2_\ep(\hat N_0)\ra N_0$, 
which contradicts Eq. (130). The cause of this contradiction is that the spatial
integrations, in the process of passing from the spatial representation to Eq. 
(131), have been done without a regularization function. To correct this defect,
one has either to return to the spatial representation and to proceed as has 
been explained above or to invoke the following argumentation.

The Bogolubov replacement $a_0\ra\sqrt{N_0}$ implies that the operators $a_0$ 
and $a_0^\dgr$ should commute in thermodynamic limit. More correctly,
\be
\label{134}
\lim_{N\ra\infty}\; \frac{<[a_0,a_0^\dgr]>_\ep}{N_0} = 0 \; ,
\ee
where $N_0=<\hat N_0>_\ep$ tends to infinity proportionally to $N$, as 
$N\ra\infty$. To correctly define $<\hat N_0^2>_\ep$, one has to consider the 
reduced quantity $<\hat N_0^2>_\ep/N_0^2$. Then, form (132) gives
$$
\lim_{N\ra\infty} \; \frac{<\hat N_0^2>_\ep}{N_0^2} =  1 \; .
$$
Form (133) yields the same answer
$$
\lim_{N\ra\infty} \; \frac{<\hat N_0^2>_\ep}{N_0^2} =
\lim_{N\ra\infty} \left ( 1 +\frac{1}{N_0}\right ) = 1 \; .
$$
In that sense, one always has
\be
\label{135}
< \hat N_0^2>_\ep \simeq N_0^2 \qquad (N\ra \infty) \; ,
\ee
independently of the form involved. Thus one always returns to Eq. (130) 
telling us that the condensate fluctuations are {\it thermodynamically 
negligible}.

One could come to catastrophic condensate fluctuations only by means of 
incorrect calculations, assuming the existence of BEC but without breaking 
gauge symmetry. In the latter case, using the Wick theorem, one would get
$$
< \hat N_0^2> \; = \; < a_0^\dgr a_0> 
( 2+ < a_0^\dgr a_0>) \; ,
$$
from where
$$
<\hat N_0^2> - < \hat N_0 >^2 = N_0(1+ N_0) \; ,
$$
with $N_0\equiv<a_0^\dgr a_0>\propto N$. As a result, one finds
$$
< \hat N_0^2> - < \hat N_0>^2 \; \propto \; N^2 \; .
$$
This what one calls the catastrophic condensate fluctuations, since the normal 
dispersion must be proportional to $N$. One often blames the grand ensemble 
for this unreasonable behavior of the dispersion. However, as has been just 
explained, it is not the grand ensemble to blame, but this fictitious 
catastrophe is merely provoked by wrong calculations. The correct way of 
treating the problem results in thermodynamically negligible condensate 
fluctuations in the sense of Eq. (130). It is worth stressing that result 
(130) is general and is valid for arbitrary Bose systems, whether uniform 
or nonuniform and even whether they are equilibrium or nonequilibrium.

\section{Representative statistical ensembles}

The basic point, analyzed and explained in this paper, is that {\it 
spontaneous gauge symmetry breaking is the necessary and sufficient 
condition for the existence of BEC}. Although it is, probably, 
possible to calculate some particular quantities of a Bose-condensed 
system without explicit symmetry breaking, but, in general, such 
a way of calculation is not correct and can lead to wrong results. 
There are two methods of gauge symmetry breaking, by introducing into 
the Hamiltonian infinitesimal sources or by means of the Bogolubov 
operator shift. These methods are equivalent in thermodynamic limit. 
However, before the latter, the methods are not equivalent, since 
they involve different operator variables possessing different 
commutation relations. To formulate this difference in a precise form, 
it is convenient to invoke the notion of representative statistical 
ensembles [31,46,48--50].

A {\it statistical ensemble} is a pair $\{\cF,\hat\rho\}$ of the 
space of microstates $\cF$ and a statistical operator $\hat\rho$. 
The latter, for an equilibrium system, is given by the Gibbs form 
characterized by the appropriate Hamiltonian, and for a nonequilibrium 
system, $\hat\rho$ is described by the Liouville equation, whose 
dynamics is governed by the related Hamiltonian.

A {\it representative statistical ensemble} is an ensemble equipped 
with all additional conditions that are necessary for a unique 
description of the given physical system [31,46]. Additional 
conditions may include symmetry properties, normalization conditions, 
conservation conditions, and so on. Actually, only representative 
ensembles have sense.

Suppose, we are considering a Bose system in the frame of the field-operator 
technique. With a field operator $\psi$, the space of microstates is the Fock 
space $\cF(\psi)$ generated by $\psi^\dgr$. The field operators, as is well 
known, do not conserve the number of particles. Therefore, working with the 
field-operator technique, it is necessary to use the grand canonical ensemble 
fixing the average number of particles $N=<\hat N>$. The statistical operator 
$\hat\rho$ is defined by a Hamiltonian $H[\psi]$ which is a functional of 
$\psi$. To emphasize this, we shall write $\hat\rho[H(\psi)]$. Let us denote 
by $\hat H[\psi]$ the energy operator. And let us define the statistical 
ensemble
\be
\label{136}
\{ \cF(\psi),\; \hat\rho(H[\psi])\} \; , \qquad 
H[\psi]\equiv \hat H[\psi] - \mu \hat N \; .
\ee
This is the standard grand canonical ensemble with the grand Hamiltonian 
$H[\psi]$, which is assumed to be gauge invariant. Ensemble (136) is 
representative for a system without BEC. But as soon as the latter appears, 
ensemble (136) is not anymore representative, since BEC necessarily requires 
gauge symmetry breaking. Accomplishing this by means of the method of 
infinitesimal sources, adding a gauge-symmetry breaking term $\hat\Gm$, we 
have the statistical ensemble
\be
\label{137}
\{ \cF(\psi), \; \hat\rho(H_\ep[\psi]) \} \; , \qquad
H_\ep[\psi] \equiv \hat H[\psi] - \mu \hat N + \ep\hat\Gm \; .
\ee
This ensemble is representative for a system with BEC, the operator averages 
being defined as the Bogolubov quasiaverages. The number of particles is now 
given by $N=<\hat N>_\ep$. The chemical potential $\mu$ is the Lagrange 
multiplier preserving the normalization condition for the number of 
particles $N$. In ensemble (137), there is the sole field variable $\psi$ 
and the only normalization condition for $N$. Hence, just one Lagrange 
multiplier is required. The representation (17) for $\psi$, as a sum of two 
terms, is rather formal, since neither of these terms comprises an 
independent variable. The parts of sum (17) are interconnected through their 
commutation relations (19), (20), and (21). It is, of course, possible to 
work with ensemble (137), invoking perturbation theory in powers of particle 
interactions, as has been done by Belyaev [35]. This, however, has two weak 
points. First, such a perturbative way makes it possible to deal exclusively 
with weakly interacting systems. Second, this way is very cumbersome because 
of the complicated commutation relations (19) and (20).

It is tempting to resort to the Bogolubov theorem and to make the Bogolubov 
shift (34), passing to the simple commutation relation (35). However, one 
has to remember that the Bogolubov shift becomes exact only in thermodynamic 
limit, but not before it. At the same time, all calculations are usually 
accomplished before thermodynamic limit. Using the commutation relations 
(35) before this limit is incorrect, makes the theory not self-consistent, 
and leads to different paradoxes. The most dangerous such a paradox is the 
Hohenberg-Martin dilemma of conserving versus gapless theories [51]. This 
paradox has recently been resolved in Refs. [32,46,48--50,52,53].

The solution to the problem is merely the necessity of being mathematically 
accurate. Yes, we can employ the Bogolubov shift (34) and to use the simple 
commutation relations (35) even before thermodynamic limit, provided $N$ is 
large. But then it is necessary to understand that the Bogolubov shift 
introduces {\it two independent variables}, the condensate wave function 
$\eta(\br)$ and the operator of uncondensed particles $\psi_1(\br)$. So, 
there are now {\it two normalization conditions}, for the number of 
condensed particles
\be
\label{138}
N_0 = \int |\eta(\br)|^2 \; d\br
\ee
and for the number of uncondensed particles
\be
\label{139}
N_1 \; = \; <\hat N_1>_\eta \; ,
\ee
where $\hat N_1$ is given by Eq. (109) and the average is defined in Eq. 
(37). Preserving two normalization conditions requires {\it two Lagrange 
multipliers}. In addition, one must be aware of the fact that the Bogolubov 
shift (34) induces a new operator representation, which is nonequivalent to 
that used in ensemble (137). Now, the space of microstates is the Fock space 
$\cF(\psi_1)$ generated by $\psi_1^\dgr$. The space $\cF(\psi_1)$ is 
asymptotically orthogonal to the space $\cF(\psi)$ of ensemble (137), with 
all operators forming the new representation on this space [32]. After 
accomplishing the Bogolubov shift, we come to the representative ensemble
\be
\label{140}
\{ \cF(\psi_1),\; \hat\rho(H[\eta,\psi_1]) \} \; , \qquad
H[\eta,\psi_1] \equiv \hat H[\eta,\psi_1] - \mu_0 N_0 -
\mu_1 \hat N_1 \; ,
\ee
with the statistical operator (38) and the grand Hamiltonian 
$H[\eta,\psi_1]$ which is a functional of two variables and which contains 
two Lagrange multipliers, $\mu_0$ and $\mu_1$, guaranteeing the validity of 
two normalization conditions (138) and (139). Since the Bogolubov shift 
explicitly breaks gauge symmetry, we do not need to invoke infinitesimal 
sources. The representative ensemble (140) is the basis for developing a 
completely self-consistent theory of Bose-condensed systems [46,48--50], 
which is free of paradoxes and which enables the description of strongly 
interacting systems with BEC [52,53].

The case of Bose-condensed systems serves as a very illustrative example 
emphasizing the principal importance of defining a representative 
statistical ensemble for the system under consideration. It may happen that 
several ensembles can correctly describe a statistical system. Then one says 
that these ensembles are equivalent. For instance, the majority of statistical 
systems can be equivalently treated by means of either grand canonical, 
canonical, or microcanonical ensembles. But this is not compulsory. Even more, 
pronouncing just the term of an ensemble, say, grand canonical, is not 
sufficient, since, as has been shown above, there can exist several grand 
canonical ensembles. Thus ensembles (136), (137), and (140) all are grand 
canonical, though they are different and not all equivalent. The most important 
is that the employed ensemble be representative. In the other case, it is unable
to correctly represent the considered system. Several ensembles are equivalent 
if and only if they are representative [46]. There are some models with 
long-range interactions for which the microcanonical and canonical ensembles 
are not equivalent [54,55]. But this in no way is a fact that would shake the 
basis of statistical mechanics. This just means that one or both of the used 
ensembles are not representative. Say, the microcanonical ensemble may occur 
to be not representative for such mean-field models. Or it may be that some 
additional constraints are to be imposed in order to correctly define the model 
and to make ensembles representative, hence, equivalent.

In conclusion, it is worth emphasizing that the principal problems 
analyzed in the present paper concern any kind of Bose systems with BEC, 
whether uniform or nonuniform. These could be uniform Bose gases and 
superfluid liquids or nonuniform clouds of trapped Bose atoms. As is 
mentioned in the Introduction, multiquark clusters can also form BEC in 
the interiors of neutrons stars and in fireballs created in heavy-ion 
collisions [17--19]. Another example is the BEC of excitons [56,57]. 
Nonuniform condensates can appear in a uniform system, when the spectrum 
of collective excitations [58,59] or the single-particle spectrum [60--62] 
touch zero at the finite momentum $\bk_0$. Then a condensate with a 
periodic structure, prescribed by the momentum $\bk_0$, arises, forming a 
coherent crystal [58,59]. A similar periodic density evolves in nuclear 
matter under pion condensation [63-72]. Periodic condensates can occur, in 
general, for any Bose particles in external fields [71,73]. Finiteness of 
a sample also can impose boundary conditions that would induce a periodic 
BEC, such as happens in magnetic films under magnon condensation [74,75]. 
The abundance of different types of BEC makes crucial the necessity of 
understanding the basic features of its correct theoretical description.

\vskip 5mm

{\bf Acknowledgement}. I am grateful to E. Lieb for useful correspondence on 
the relation between BEC and gauge symmetry breaking.

\end{document}